\documentclass[journal]{IEEEtran}

\usepackage{cite}
\usepackage{mathrsfs}
\usepackage{amsfonts}
\usepackage{ifpdf}
\usepackage{stfloats}
\ifCLASSINFOpdf
\usepackage[pdftex]{graphicx}
\else \fi
\usepackage[cmex10]{amsmath}
\usepackage{array}
\usepackage{mdwmath}
\usepackage{mdwtab}
\usepackage{eqparbox}
\usepackage[tight,footnotesize]{subfigure}
\usepackage{algorithmic}
\usepackage{algorithm}
\usepackage{amsmath}
\usepackage{epsfig}
\usepackage{ae}
\usepackage{amssymb}
\usepackage{times}
\usepackage{amsmath}   
\usepackage{booktabs}  
\usepackage{threeparttable} 
\usepackage{multirow}       
\usepackage{xcolor}
\usepackage{paralist}
\usepackage{cases}
\newtheorem{theorem}{Theorem}
\newtheorem{lemma}{Lemma}

\newtheorem{remark}{Remark}

\ifCLASSINFOpdf

\else

\fi

\begin{document}

\title{Physical Layer Security-Aware Routing and Performance Tradeoffs in Ad Hoc Networks}

\author{
Yang Xu$^{1}$, Jia Liu$^{2}$, Yulong Shen$^{3}$, Xiaohong Jiang$^{4}$ and Norio Shiratori$^{5}$
\thanks{$^1$Y.~Xu is with the School of Economics and Management, Xidian University, Xi'an 710071, China. Email: yxu@xidian.edu.cn}
\thanks{$^2$J.~Liu is with the Cyber Security Research Center, National Institute of Informatics, Tokyo 101-8430, Japan, and also with the State Key Laboratory of Integrated Services Network, Xidian University, Xi'an 710071, China. Email: jliu@nii.ac.jp,}
\thanks{$^3$Y.~Shen is with the State Key Laboratory of Integrated Services Network, Xidian University, Xi'an 710071, China. Email: ylshen@mail.xidian.edu.cn}
\thanks{$^4$X.~Jiang is with the School of Systems Information Science, Future University Hakodate, Hakodate 041-8655, Japan. Email: jiang@fun.ac.jp}
\thanks{$^5$N. Shiratori is with the Research and Development Initiative, Chuo University, Tokyo 112-8551, Japan. Email: norio@shiratori.riec.tohoku.ac.jp}
}

\maketitle

\begin{abstract}
The application of physical layer security in ad hoc networks has attracted considerable academic attention recently. However, the available studies mainly focus on the single-hop and two-hop network scenarios, and the price in terms of degradation of communication quality of service (QoS) caused by improving security is largely uninvestigated. As a step to address these issues, this paper explores the physical layer security-aware routing and performance tradeoffs in a multi-hop ad hoc network. Specifically, for any given end-to-end path we first derive its connection outage probability (COP) and secrecy outage probability (SOP) in closed-form, which serve as the performance metrics of communication QoS and transmission security, respectively. Based on the closed-form expressions, we then study the security-QoS tradeoffs to minimize COP (resp. SOP) conditioned on that SOP (resp. COP) is guaranteed. With the help of analysis of a given path, we further propose the routing algorithms which can achieve the optimal performance tradeoffs for any pair of source and destination nodes in a distributed manner. Finally, simulation and numerical results are presented to validate the efficiency of our theoretical analysis, as well as to illustrate the security-QoS tradeoffs and the routing performance.
\end{abstract}

\begin{IEEEkeywords}
Ad hoc networks, physical layer security, QoS, routing.
\end{IEEEkeywords}

\IEEEpeerreviewmaketitle

\section{Introduction}

\subsection{Background and Related Works}
The ad hoc network represents a class of self-organizing network architecture, which consists of nodes communicating with each other over peer-to-peer wireless channels without centralized infrastructure \cite{Perkins_BOOK08}. Since ad hoc networks can be flexibly deployed and reconfigured at very low cost, they are highly promising for many critical applications, such as disaster relief, emergency rescue, daily information exchange, traffic off-loading and coverage extension for 5G cellular networks \cite{Ramanathan_CM02,Tehrani_magzine14}. To facilitate the application and commercialization of ad hoc networks, protecting their transmission security is of great significance \cite{Kaufman_BOOK02}. However, due to the broadcast nature of wireless channel and the lack of central administration, it is very challenging for the traditional cryptographic-based security techniques \cite{Kahate_BOOK13} to be applied in such a distributed ad hoc network.

As a complementary technique of cryptographic-based methods, physical layer security, an information-theoretic approach which exploits the fundamental characteristics of wireless channel to achieve perfect secrecy, has been extensively studied over the past few decades. Based on the results of Shannon in \cite{Shannon_BST49}, Wyner first indicated that perfect secrecy is achievable when the condition of main channel between transmitter and receiver is better than that of wiretap channel between transmitter and eavesdropper \cite{Wyner_BS75}. Following this line, many research activities have been devoted to the study of physical layer security under various channel models, such as the broadcast channel \cite{Csiszar_IT78}, Gaussian wiretap channel \cite{Cheong_IT78}, two-way wiretap channel \cite{Tekin_IT08}, multi-access channel \cite{Liang_IT08} and MIMO wiretap channel \cite{Oggier_IT11}. Meanwhile, diverse approaches for improving physical layer security have been proposed in the literature. The works of \cite{Lai_IT08,Krikidis_TWC09,Dong_TSP10,Vilela_IFS11,Mo_COMML12,Hui_SPL15,Tang_COMML15} demonstrated that the strategies of cooperative jamming and relay selection can be utilized to enhance physical layer security. The works of \cite{Harrison_SPM13,He_IT14} indicated that physical layer security can also be facilitated by applying coding schemes. Moreover, the combinations of physical layer security with other techniques such as power allocation, signal processing, and cross-layer optimization were explored in \cite{Jeong_TSP11}, \cite{Hong_SPM13} and \cite{Zhou_CM15}, respectively.

Since physical layer security has the advantage of low computational complexity and can be easily implemented in a distributed manner, its application in ad hoc networks has attracted considerable academic attention recently \cite{vasudevan_MOB10,Zhang_TON11,Zhou_TWC11,Goeckel_JSAC11,Koyluoglu_IT12,Zou_JSAC13,Xie_IT14,Duy_IET15,Karas_CL16}. For a large-scale ad hoc network, Vasudevan \emph{et al.} \cite{vasudevan_MOB10} investigated the asymptotic behaviors of security-capacity tradeoff as the number of network nodes tends to infinity. The price in terms of performance degradation for ensuring physical layer security in ad hoc networks was explored under the large-scale network scenario  \cite{Zhang_TON11} and single-hop network scenario \cite{Zhou_TWC11}, respectively. Goeckel \emph{et al.} \cite{Goeckel_JSAC11} indicated that the artificial noise generated by cooperative relays can be utilized to achieve everlasting secrecy in a two-hop ad hoc network. Koyluoglu \emph{et al.} \cite{Koyluoglu_IT12} studied the scaling behaviors of ad hoc networks under secrecy constraints. They demonstrated that under the path loss model, a secure rate of $\Omega(\frac{1}{\sqrt{n}})$ is achievable if the density of eavesdropper is below some threshold\footnote{$n$ is the number of source-destination pairs and please kindly refer to \cite{Cormen_BOOK01} for the asymptotic notations.}; while under the ergodic fading model, a constant secret rate can be achieved for sufficiently large $n$. For a two-hop relay ad hoc network, Zou \emph{et al.} \cite{Zou_JSAC13} explored the cooperative-based relay selection schemes to improve transmission security against eavesdropping attack. Xie and Ulukus \cite{Xie_IT14} considered the single-hop ad hoc network with four fundamental wireless channels, and studied its secure degrees of freedom as well as provided the corresponding achievable schemes. Duy \emph{et al.} \cite{Duy_IET15} evaluated the secrecy performance of a two-hop cooperative relay network under the impact of co-channel interference and proposed an optimal relay selection scheme to maximize the secrecy capacity. Karas \emph{et al.} \cite{Karas_CL16} derived the closed-form expressions for SOP in a single-hop cellular system with the consideration of eavesdropper's location uncertainty. For a detailed survey on physical layer security and its applications in ad hoc networks, please kindly refer to \cite{Mukherjee_CST14} and references therein.

\subsection{Motivation and Our Contributions}
Although there have been extensive works for studying physical layer security in wireless networks, they mainly focus on either the single-hop and two-hop network scenarios, or the asymptotic large-scale network scenarios, while the research of physical layer security in multi-hop ad hoc networks which fills the significant gap between those two extremes is largely untouched and thus remains a technical challenge. By now, some initial results have been reported on the study of physical layer security in multi-hop ad hoc networks \cite{Saad_TWC12,Ghaderi_TMC15,Yao_TCOM16,Lee_TWC16}. Specifically, Saad \emph{et al.} \cite{Saad_TWC12} proposed a tree-formation game to choose secure paths in multi-hop ad hoc networks. Later, Ghaderi \emph{et al.} \cite {Ghaderi_TMC15} explored the minimum energy routing which can guarantee security for multi-hop ad hoc networks. More recently, Yao \emph{et al.} \cite{Yao_TCOM16} studied the physical layer security-based routing in multi-hop ad hoc networks with decode-and-forward relaying, and Lee \cite{Lee_TWC16} proposed an optimal power allocation strategy for maximizing the secrecy rate in a special multi-hop relay network with single source-destination pair. 

It is notable that security usually comes with a cost in terms of performance degradation \cite{Zhang_TON11,Zhou_TWC11}, thus the tradeoffs between security and other network performance should be carefully addressed for a practical multi-hop ad hoc network. In \cite{Taleb_GC10,Taleb_SCN12} and our previous work \cite{Xu_ICC16}, the issue of integrating security and quality of service (QoS) under some network scenarios was investigated. In \cite{Zou_TVT14,Zou_TCOM15}, the security-reliability tradeoff was explored in cooperative relay networks and cognitive radio systems, respectively, where the two-hop network scenarios were considered and corresponding optimal relay selection schemes were proposed. While in this paper, for the first time, we explore the tradeoffs between transmission security and communication QoS in a multi-hop ad hoc network. We consider a general multi-hop ad hoc network with randomly distributed legitimate nodes, cooperative jammers and malicious eavesdroppers, and analyze the connection outage probability (COP) and secrecy outage probability (SOP) for a given path. Based on the outage probability analysis, we study the COP and SOP tradeoffs, and further propose the routing algorithms which can achieve the optimal performance with the guaranteed communication QoS and transmission security in the concerned ad hoc network.

The main contributions of this paper are summarized as follows:

\begin{itemize}
\item
For any given end-to-end path in a general multi-hop ad hoc network where jammers and malicious eavesdroppers are randomly distributed following the independent Poisson point processes, we derive its COP and SOP in closed-form, which serve as the performance metrics of communication QoS and transmission security, respectively.

\item
We formulate the security-QoS tradeoffs of a given path as two constrained optimization problems and provide corresponding analysis to obtain the optimal solutions. Based on the results of a given path, we further propose the routing algorithms which can find the optimal path between any pair of source and destination nodes, and allocate transmission power for each node on the path to achieve the optimal performance.

\item
We provide extensive simulation and numerical results to validate the efficiency of our theoretical analysis, and illustrate the security-QoS tradeoffs as well as the performance of proposed routing algorithms.
\end{itemize}

\subsection{Paper Organization}
The remainder of this paper is organized as follows. Section~\ref{section:preliminaries} introduces the preliminaries involved in this paper. The expressions of COP and SOP are derived in Section~\ref{section:outage_probabilities}. We explore the tradeoffs between COP and SOP in Section~\ref{section:optimization} and propose the routing algorithms in Section~\ref{section:routing}. Finally, Section~\ref{section:numerical_results} presents the simulation and numerical results, and Section~\ref{section:conclusion} concludes this paper.

\section{Preliminaries} \label{section:preliminaries}
In this section, we introduce the network model, wireless channel model and performance metrics involved in this study.

\subsection{Network Model}
Following the network model in \cite{Ghaderi_TMC15}, we consider a general multi-hop ad hoc network which consists of arbitrarily distributed legitimate nodes, cooperative jammers and malicious eavesdroppers. A $K$-hop path (route) $\Pi=\left\langle l_{1},\ldots,l_{K}\right\rangle$ in the network is formed by $K$ links from $l_{1}$ to $l_{K}$, and a link $l_k \in \Pi$ connects two legitimate nodes $S_k$ and $D_k$ on path $\Pi$. We assume that each link $l_k$ is exposed to a set of eavesdroppers denoted by $\Phi_{E}=\left\{E_i,i=1,2,\ldots\right\}$. The locations of eavesdroppers are unknown since they usually work in a passive way. In order to statistically evaluate the network performance, an independent homogeneous Poisson point process (PPP) \cite{Zhou_TWC11,Ghaderi_TMC15,Yao_TCOM16} with density $\lambda_E$ is applied to characterize the distributions of eavesdroppers' locations. Furthermore, we assume that whenever a legitimate node $S_k$ transmits a message, a set of external nodes $\Phi_{J}=\left\{J_j,j=1,2,\ldots\right\}$, called jammers, cooperate with the legitimate node by jamming the message at eavesdroppers. The locations of jammers also follow an independent homogeneous PPP with density $\lambda_J$ and the transmission power of each jammer is the same, denoted by $P_{\bar{J}}$.

\subsection{Wireless Channel Model}
We consider the decode-and-forward (DF) relaying scheme and assume that the instantaneous wireless channel state between any pair of nodes is unavailable but can be statistically characterized by the large-scale path loss along with the small-scale Rayleigh fading \cite{Zou_JSAC13,Ghaderi_TMC15,Kim_IT09,Rappaport_BOOK96}. We also apply the randomization strategy widely used in other works \cite{Mo_COMML12,Koyluoglu_IT12,Ghaderi_TMC15}, with which each link transmits independent randomization signal such that eavesdroppers cannot use combining techniques to combine the received signals from multiple hops. In addition, we assume that the network is interference-limited and thus the noise at the receiver is negligible \cite{Zhou_TWC11}. The reason of adopting interference-limited assumption is that the mathematical tractability of this assumption allows us to gain important insights into the security-QoS tradeoffs and routing protocol design in ad hoc networks, as shown in later context. More formally, regarding a transmission from node $S$ to node $D$, let $P_S$ denote the transmission power of $S$, $x_S$ and $x_{J_j}$ denote the normalized (unit power) symbol stream to be transmitted by $S$ and its $j^{th}$ jammer $J_j$, respectively, and $y_D$ denote the received signal at $D$. Then $y_D$ can be expressed as\footnote{We don't consider mutual interference between legitimate links since it can be avoided by mature media access control techniques.}:
\begin{equation}
y_D=\frac{\sqrt{P_S} h_{S,D}}{d_{S,D}^{\alpha/2}} x_S +  \sum_{J_j \in \Phi_J}\frac{ \sqrt{P_{\bar{J}}} h_{J_j,D}}{d_{J_j,D}^{\alpha/2}}x_{J_j},  \label{eq:signal_received_at_D}
\end{equation}
where $d_{S,D}$ and $h_{S,D}$ (resp. $d_{J_j,D}$ and $h_{J_j,D}$) are the distance and the fading coefficient of wireless channel between $S$ (resp. $J_j$) and $D$, $\alpha$ is the path-loss exponent (typically between 2 and 6), $|h_{S,D}|^2$ (resp. $|h_{J_j,D}|^2$) is exponentially distributed with $\mathbb{E} \{|h_{S,D}|^2\}=1$ (resp. $\mathbb{E} \{|h_{J_j,D}|^2\}=1$). Similarly, for an eavesdropper $E \in \Phi_E$, the signal $y_E$ received at $E$ is given by
\begin{equation}
y_E=\frac{\sqrt{P_S}h_{S, E}}{d_{S, E}^{\alpha/2}} x_S + \sum_{J_j \in \Phi_J} \frac{\sqrt{P_{\bar{J}}} h_{J_j,E}} {d_{J_j, E}^{\alpha/2}} x_{J_j},  \label{eq:signal_received_at_E}
\end{equation}
where $d_{S,E}$ and $h_{S,E}$ (resp. $d_{J_j,E}$ and $h_{J_j,E}$) are the distance and the fading coefficient of wiretap link between $S$ (resp. $J_j$) and $E$, $|h_{S,E}|^2$ (resp. $|h_{J_j,E}|^2$) is exponentially distributed with $\mathbb{E} \{|h_{S,E}|^2\}=1$ (resp. $\mathbb{E} \{|h_{J_j,E}|^2\}=1$).

\subsection{Performance Metrics}

Following the definitions in \cite{Zhou_TWC11,Ghaderi_TMC15}, the performance metrics involved in this paper are defined as follows:

\textbf{Connection Outage Probability}: The event of \emph{connection outage} refers to the case when the signal-to-interference ratio (SIR) at the intended receiver is below a required threshold $\gamma_C$, such that the message cannot be correctly decoded by the receiver. The \emph{connection outage probability} (COP) $P_{co}$ is defined as the probability that the event of \emph{connection outage} happens.

\textbf{Secrecy Outage Probability}: The event of \emph{secrecy outage} refers to the case when the SIR at one or more eavesdroppers is above a required threshold $\gamma_E$, such that the message can be decoded by the eavesdropper(s). The \emph{secrecy outage probability} (SOP) $P_{so}$ is defined as the probability that the event of \emph{secrecy outage} happens.

Notice that in Wyner's encoding scheme \cite{Wyner_BS75,Zhou_TWC11}, the transmitter chooses two rates, the rate of transmitted codewords $R_t$ and the rate of the confidential messages $R_s$. The rate difference $R_e=R_t-R_s$ reflects the cost of securing the messages against the eavesdroppers. If the legitimate channel capacity is less than $R_t$, the connection outage happens. While if the wiretap channel capacity is higher than $R_e$, the secrecy outage happens. According to Shannon's Theorem, the channel capacity is determined by the corresponding SIR at the receiver. Thus, our performance metrics can be easily mapped to those based on Wyner's encoding scheme \cite{Wyner_BS75,Krikidis_TWC09,Mo_COMML12}, where the conversions between the SIR thresholds and the code rates are $\gamma_C=2^{R_t}-1$ and $\gamma_E=2^{R_e}-1$, and the results in this paper also applies to the Wyner's encoding scheme. 

\begin{remark}
The performance metrics COP and SOP are equivalent to the metrics outage probability (OP) and intercept probability (IP) defined in \cite{Zou_TVT14,Zou_TCOM15} by applying Shannon's Theorem, respectively. Following the statement in \cite{Zou_TVT14,Zou_TCOM15}, COP and SOP are of high significance as COP represents the communication QoS of a network user, while SOP serves as a measure of the transmission security level.
\end{remark}

\section{Outage Probabilities Analysis} \label{section:outage_probabilities}

In this section, we derive the exact expressions of COP and SOP for a given path, which will help us explore the performance tradeoffs in Section~\ref{section:optimization}.

\subsection{COP Analysis}
Regarding the COP of a given path, we have the following lemma.

\begin{lemma}
For a concerned ad hoc network with the network model and wireless channel model as described in Section~\ref{section:preliminaries}, the COP of a $K$-hop path $\Pi=\left\langle l_{1},\ldots,l_{K}\right\rangle$ is given by
\begin{equation}
P_{co}(\Pi)=1-\exp\left(-A_{co}\sum_{l_k \in \Pi} d_{S_k, D_k}^2 P_{S_k}^{-\frac{2}{\alpha}}\right), \label{eq:COP}
\end{equation}
where $A_{co} =\lambda_J \pi \left(\gamma_C P_{\bar{J}}\right)^{\frac{2}{\alpha}} \Gamma(1-\frac{2}{\alpha})\Gamma(1+\frac{2}{\alpha})$, $\Gamma(\cdot)$ is a gamma function, $P_{S_k}$ denote the transmission power of $S_k$.
\end{lemma}

\begin{IEEEproof}
We first derive the COP for a link $l_k$ on path $\Pi$, which is termed as $P_{co}(l_k)$. Based on the wireless channel model of Expression (\ref{eq:signal_received_at_D}) and the definition of COP, $P_{co}(l_k)$ can be determined as:
\begin{equation}
P_{co}(l_k) = \mathbb{P} \left\{\frac{P_{S_k} |h_{{S_k},{D_k}}|^2/d_{S_k,D_k}^\alpha}{\sum_{J_j \in \Phi_J} P_{\bar{J}}  |h_{J_j,D_k}|^2/d_{J_j,D_k}^\alpha} < \gamma_C \right\}, \label{eq:pco_lk}
\end{equation}
which can be further rewritten as (\ref{eq:pco_lk_1}), shown at the top of the next page.
\begin{figure*}[!t]
\centering
\begin{align}
P_{co}(l_k) & = 1-\mathbb{E}_{\Phi_J}\left\{\mathbb{E}_{h_{J_j,D_k}} \left\{\exp\left(\frac {-\gamma_C \sum_{J_j \in \Phi_J} P_{\bar{J}} |h_{J_j,D_k}|^2/d_{J_j,D_k}^\alpha}{P_{S_k}/d_{S_k,D_k}^\alpha}\right)\right\} \right\} \nonumber\\
            & = 1-\mathbb{E}_{\Phi_J}\left\{\prod\limits_{J_j \in \Phi_J}\mathbb{E}_{h_{J_j,D_k}} \left\{\exp\left(\frac {-\gamma_C P_{\bar{J}} |h_{J_j,D_k}|^2/d_{J_j,D_k}^\alpha}{P_{S_k}/d_{S_k,D_k}^\alpha}\right)\right\} \right\} \nonumber\\
            & = 1-\mathbb{E}_{\Phi_J}\left\{\prod\limits_{J_j \in \Phi_J}\left\{\int_0^\infty \exp\left[-\left(\frac {\gamma_C P_{\bar{J}} /d_{J_j,D}^\alpha}{P_{S_k}/d_{S_k,D_k}^\alpha}+1\right)x\right] dx\right\}\right\} \nonumber\\
          &= 1-\mathbb{E}_{\Phi_J}\left\{\prod\limits_{J_j \in \Phi_J}\frac {1}{1+\frac {\gamma_C P_{\bar{J}}/d_{J_j,D_k}^\alpha}{P_{S_k}/d_{S_k,D_k}^\alpha}}\right\}. \label{eq:pco_lk_1}
\end{align}
\hrulefill
\end{figure*}

Notice that for a homogeneous PPP, the corresponding probability generating functional (PGFL) is given by \cite{Weber_TCOM10}
\begin{align}
\mathbb{E}_{\Phi_J}\left\{\prod\limits_{J_j \in \Phi_J}f(z_{J_j})\right\} &= \exp\left[-\lambda_J \int_{\mathbb{R}^2} 1-f(z_{J_j}) dz_{J_j} \right] \nonumber\\
&= \exp\left[-2\pi\lambda_J \int_{0}^{\infty} (1-f(r))rdr\right], \label{eq:PGFL}
\end{align}
where $z_{J_j}$ is the location of $J_j$. By applying PGFL in (\ref{eq:pco_lk_1}), then $P_{co}(l_k)$ can be expressed as:
\begin{align}
P_{co}(l_k) &=1-\exp\left[-2\pi \lambda_J \int_{0}^{\infty}\left( \frac {1}{1+\frac {P_{S_k}/d_{S_k,D_k}^\alpha}{\gamma_C P_{\bar{J}}/r^\alpha}}\right)rdr\right]  \nonumber\\
&=1- \exp\left(-A_{co}  d_{S_k, D_k}^2 P_{S_k}^{-\frac{2}{\alpha}}\right). \label{eq:pco_lk_2}
\end{align}

Due to the randomization strategy, each legitimate receiver can only decode the signal of each hop individually according to the link SIR. Therefore, based on the COP of a link $l_k$, the COP $P_{co}(\Pi)$ of the $K$-hop path $\Pi$ can be finally determined as:
\begin{align}
P_{co}(\Pi)& =1-\prod_{l_k \in \Pi} \left[1-P_{co}(l_k)\right]  \label{eq:E2E_COP} \\
           & =1-\exp\left(-A_{co}\sum_{l_k \in \Pi} d_{S_k, D_k}^2 P_{S_k}^{-\frac {2}{\alpha}}\right). \nonumber
\end{align}
\end{IEEEproof}

We can see from Formula~(\ref{eq:COP}) that $P_{co}(\Pi)$ is an increasing function of $\lambda_J$, $P_{\bar{J}}$ and $\gamma_C$, while being a decreasing function of $P_{S_k}$.

\subsection{SOP Analysis}
Regarding the SOP of a given path, we have the following lemma.

\begin{lemma}
For a concerned ad hoc network with the network model and wireless channel model as described in Section~\ref{section:preliminaries}, the SOP of a $K$-hop path $\Pi=\left\langle l_{1},\ldots,l_{K}\right\rangle$ is given by
\begin{equation}
P_{so}(\Pi)=1-\exp\left(-B_{so}\sum_{l_k \in \Pi} P_{S_k}^{\frac{2}{\alpha}}\right), \label{eq:SOP}
\end{equation}
where $\displaystyle B_{so}=\frac{\lambda_E}{\lambda_J} \left[ \left(\gamma_E P_{\bar{J}}\right)^{\frac {2}{\alpha}} \Gamma(1-\frac {2}{\alpha}) \Gamma(1+\frac {2}{\alpha}) \right]^{-1} $.
\end{lemma}

\begin{IEEEproof}
We first derive the SOP for a link $l_k$ on path $\Pi$, which is termed as $P_{so}(l_k)$. Based on the wireless channel model of Expression (\ref{eq:signal_received_at_E}) and the definition of SOP, $P_{so}(l_k)$ can be determined as (\ref{eq:pso_lk}), shown at the top of the next page.
\begin{figure*}[!t]
\centering
\begin{equation}
P_{so}(l_k)  \!=\! 1\!-\!\mathbb{E}_{\Phi_J}\left\{\mathbb{E}_{\Phi_E}\left\{\prod\limits_{E_i \in \Phi_E}\left\{1\!-\!\mathbb{P} \left\{ \left. \frac {P_{S_k} |h_{{S_k},{E_i}}|^2/d_{S_k,E_i}^\alpha}{\sum\limits_{J_j \in \Phi_J} P_{\bar{J}}  |h_{J_j, E_i}|^2/d_{J_j, E_i}^\alpha} \!>\! \gamma_E \right|{\Phi_E,\Phi_J} \right\} \right\} \right\} \right\}.  \label{eq:pso_lk}
\end{equation}
\hrulefill
\end{figure*}
Applying the PGFL technique for the PPP $\Phi_E$, then Equation (\ref{eq:pso_lk}) can be re-expressed as (\ref{eq:pso_lk_5}), shown at the top of the next page,
\begin{figure*}[!t]
\centering
\begin{align}
P_{so}(l_k) & = 1-\mathbb{E}_{\Phi_J}\left\{\exp\left\{-\lambda_E \int_{\mathbb{R}^2}\mathbb{P} \left\{ \left. \frac {P_{S_k} |h_{{S_k},{E_i}}|^2/d_{S_k,E_i}^\alpha}{\sum_{J_j \in \Phi_J} P_{\bar{J}} |h_{J_j,{E_i}}|^2/d_{J_j, E_i}^\alpha} > \gamma_E \right|{\Phi_J} \right\}dz_{E_i}\right\} \right\} \label{eq:pso_lk_2} \\
            & \leq 1-\exp\left\{-\lambda_E \int_{\mathbb{R}^2} \mathbb{P} \left\{ \frac {P_{S_k} |h_{{S_k},{E_i}}|^2/d_{S_k,E_i}^\alpha}{\sum_{J_j \in \Phi_J} P_{\bar{J}}  |h_{J_j,{E_i}}|^2/d_{J_j, E_i}^\alpha} > \gamma_E  \right\}dz_{E_i}\right\} \label{eq:pso_lk_3} \\
            & = 1-\exp\left\{-\lambda_E \int_{\mathbb{R}^2}\exp\left[-\lambda_J \pi  d_{S_k, E_i}^2 \left(\gamma_E \frac{P_{\bar{J}}}{P_{S_k}}\right)^{\frac {2}{\alpha}}\Gamma(1-\frac {2}{\alpha})\Gamma(1+\frac {2}{\alpha})\right]dz_{E_i}\right\}  \label{eq:pso_lk_4} \\
            & = 1-\exp\left\{ -2\pi \lambda_E \int_{0}^{\infty} \exp\left[-\lambda_J \pi r^2 \left(\gamma_E \frac{P_{\bar{J}}}{P_{S_k}}\right)^{\frac {2}{\alpha}}\Gamma(1-\frac {2}{\alpha})\Gamma(1+\frac {2}{\alpha})\right]rdr \right\} \nonumber \\
            & = 1-\exp\left(-B_{so} P_{S_k}^{\frac{2}{\alpha}}\right), \label{eq:pso_lk_5}
\end{align}
\hrulefill
\end{figure*}
where (\ref{eq:pso_lk_3}) follows from the Jensen's inequality, and (\ref{eq:pso_lk_4}) follows from the same procedures which transform (\ref{eq:pco_lk}) into (\ref{eq:pco_lk_2}).

Due to the randomization strategy, each eavesdropper can only decode the signal of each hop individually according to the eavesdropping link SIR. Therefore, based on the SOP of a link $l_k$, the SOP $P_{so}(\Pi)$ of the $K$-hop path $\Pi$ can be finally determined as:
\begin{align}
P_{so}(\Pi)&=1-\prod_{l_k \in \Pi}\left[ 1-P_{so}(l_k)\right] \label{eq:E2E_SOP}\\
           &=1-\exp\left(-B_{so}\sum_{l_k \in \Pi} P_{S_k}^{\frac {2}{\alpha}}\right). \nonumber
\end{align}
\end{IEEEproof}

We can see from Formula~(\ref{eq:SOP}) that $P_{so}(\Pi)$ is an increasing function of $P_{S_k}$ and $\lambda_E$, while being a decreasing function of $\gamma_E$, $\lambda_J$ and $P_{\bar{J}}$. It is notable that the statistical properties of the locations of eavesdroppers and jammers as well as the corresponding channel states have been carefully incorporated into the derivations of COP and SOP.

\begin{remark}
For a given ad hoc network, the network parameters $\lambda_J$, $P_{\bar{J}}$, $\gamma_C$, $\lambda_E$ and $\gamma_E$ are usually pre-determined, the controllable parameter is the transmission power of each transmitter. It is worth noting that increasing $P_{S_k}$ will lead to a decrease in $P_{co}(\Pi)$ and an increase in $P_{so}(\Pi)$, which agrees with the intuition that a larger transmission power can bring about a larger SIR at the intended receiver to gain a lower COP, at the same time it comes with the cost of a higher SOP since there is also a larger SIR at the eavesdroppers. This observation indicates that by adjusting the transmission power of each transmitter on path $\Pi$, we can achieve performance tradeoffs between COP and SOP.
\end{remark}

Since the performance tradeoffs between COP and SOP exist, a problem of insight is how to optimize (minimize) one outage probability while ensuring that another outage probability is below some pre-specified threshold. This problem is termed as the optimal performance tradeoffs and will be analyzed in the next section.

\section{Optimal Performance Tradeoffs} \label{section:optimization}

In this section, we formally define the optimal performance tradeoffs as the problems of secure-based optimal COP (SO-COP) and QoS-based optimal SOP (QO-SOP), and provide corresponding solutions, respectively.

\subsection{SO-COP: Secure-based Optimal COP}

We first analyze how to achieve optimal QoS performance (minimal COP) conditioned on that secure performance is ensured (SOP is below some pre-specified threshold), which is termed as the problem SO-COP.

Let $\beta_{so}$ ($0<\beta_{so}<1$) denote the pre-specified constraint on SOP of path $\Pi$, then the problem SO-COP can be formally defined as the following optimization issue:
\begin{eqnarray}
&\mathop {\min }\limits_{l_k \in \Pi,  P_{S_k} } & P_{co}(\Pi) \label{eq:SO-COP} \\
&s.t. & P_{so}(\Pi)\leq \beta_{so}.  \label{eq:st-COP}
\end{eqnarray}

Regarding the problem SO-COP (\ref{eq:SO-COP})-(\ref{eq:st-COP}), we have the following theorem.

\begin{theorem} \label{theorem:SO-COP}
For a concerned multi-hop ad hoc network, where the densities of eavesdroppers and jammers are $\lambda_E$ and $\lambda_J$, respectively, the required SIRs for an intended receiver correctly decoding the message and an eavesdropper successfully intercepting the message are $\gamma_C$ and $\gamma_E$, respectively, the constraint on transmission security is $\beta_{so}$, then the optimal solution (i.e., optimal transmission power) of problem SO-COP is determined as:
\begin{equation}
P_{S_k}^{\text{SO-COP}}=\left( -\frac{\ln(1-\beta_{so})}{B_{so}} \cdot \frac{d_{S_k,D_k}}{\sum\limits_{l_k \in \Pi}d_{S_k,D_k}} \right)^{\alpha/2},  \quad \ l_k \in \Pi,
\label{eq:optimal_P_COP}
\end{equation}
and the optimal achievable COP with the guaranteed SOP is given by
\begin{equation}
P_{co}^*(\Pi)=1-\exp{\left[\frac{\lambda_E \pi}{\ln(1-\beta_{so})}\left(\frac{\gamma_C}{\gamma_E}\right)^{\frac {2}{\alpha}}\left(\sum\limits_{l_k \in \Pi} d_{S_k, D_k}\right)^2\right]}. \label{eq:optimal_COP}
\end{equation}
\end{theorem}

\begin{IEEEproof}
Let $F_k=P_{S_k}^{2/\alpha}$, then $P_{co}(\Pi)$ in Formula (\ref{eq:COP}) and $P_{so}(\Pi)$ in Formula (\ref{eq:SOP}) can be re-expressed as:
\begin{align}
& P_{co}(\Pi) =1- \exp\left(-A_{co} \sum\limits_{l_k \in \Pi}{\frac{d_{S_k,D_k}^2}{F_k}} \right), \label{eq:COP_1}\\
& P_{so}(\Pi) =1- \exp\left(-B_{so} \sum\limits_{l_k \in \Pi}{F_k}\right). \label{eq:SOP_1}
\end{align}
Substituting (\ref{eq:SOP_1}) into (\ref{eq:st-COP}), we have
\begin{equation}
\sum\limits_{l_k \in \Pi} {F_k} \leq -\frac{\ln(1-\beta_{so})}{B_{so}} \triangleq \epsilon_{so}. \label{eq:F_k_COP}
\end{equation}
Notice that $P_{co}(\Pi)$ in (\ref{eq:COP_1}) is a decreasing function of $F_k$ while the objective in (\ref{eq:SO-COP}) is to minimize $P_{co}(\Pi)$, so the inequality constraint (\ref{eq:F_k_COP}) can be replaced by the equality constraint $\sum\limits_{l_k \in \Pi} {F_k} = \epsilon_{so}$. Therefore, the problem SO-COP is equivalent to the following optimization issue\footnote{We can see from this optimization issue that except the transmission power $P_{S_k}$ of each link is variable, the distance $d_{S_k,D_k}$ is fixed and other quantities are known in a statistical perspective of view.}:
\begin{eqnarray}
&\mathop {\min }\limits_{l_k \in \Pi,  F_k } & \sum\limits_{l_k \in \Pi} \frac{d_{S_k,D_k}^2}{F_k} \label{eq:SO-COP_1} \\
&s.t. &  \sum\limits_{l_k \in \Pi} {F_k} = \epsilon_{so}.  \label{eq:st-COP_1}
\end{eqnarray}

To solve the above optimization issue, we apply the method of Lagrange multipliers \cite{Lange_BOOK13}. Then, we obtain the following $K$ equations:
\begin{equation}
\begin{aligned}
\frac{\partial}{\partial {F_k}}\left\{ \sum_{l_k \in \Pi} {\frac {d_{S_k, D_k}^2 }{F_k}} + \theta_1 \left( \sum_{l_k \in \Pi} F_k- \epsilon_{so} \right) \right\} \Bigg{|}_{F_k^*}=0, \\
l_k \in \Pi,
\end{aligned}
\end{equation}
where $\theta_1$ is the Lagrange multiplier, and we have
\begin{align}
&-\frac{d_{S_k, D_k}^2}{(F_k^*)^2}+\theta_1=0, \  l_k \in \Pi, \nonumber \\
\Rightarrow &F_k^*=\frac{1}{\sqrt \theta_1} d_{S_k, D_k}, \ l_k \in \Pi. \label{eq:F_k_COP_1}
\end{align}
Substituting (\ref{eq:F_k_COP_1}) into (\ref{eq:st-COP_1}), $\theta_1$ can be determined as:
\begin{equation}
\theta_1=\left(\frac{1}{\epsilon_{so}}\sum_{l_k \in \Pi} d_{S_k, D_k}\right)^2. \label{eq:theta_1}
\end{equation}
Substituting (\ref{eq:theta_1}) into (\ref{eq:F_k_COP_1}), we have
\begin{equation}
F_k^*=\epsilon_{so}\frac{d_{S_k, D_k}}{\sum\limits_{l_k \in \Pi} d_{S_k, D_k}}, \ l_k \in \Pi.
\end{equation}
Thus, the optimal transmission power $P_{S_k}^{\text{SO-COP}}$ of node $S_k$ is given by
\begin{equation*}
P_{S_k}^{\text{SO-COP}}=\left( -\frac{\ln(1-\beta_{so})}{B_{so}} \cdot \frac{d_{S_k,D_k}}{\sum\limits_{l_k \in \Pi}d_{S_k,D_k}} \right)^{\alpha/2},
\end{equation*}
and the minimum COP $P_{co}^*(\Pi)$ of path $\Pi$ under the condition that $P_{so}(\Pi) \leq \beta_{so}$ is determined as:
\begin{align*}
P_{co}^*(\Pi) &=1- \exp\left(-A_{co} \sum\limits_{l_k \in \Pi}{\frac{d_{S_k,D_k}^2}{F_k^*}} \right) \nonumber \\
           &=1-\exp\left[-A_{co} \frac{1}{\epsilon_{so}}\left(\sum\limits_{l_k \in \Pi} d_{S_k, D_k}\right)^2\right]\nonumber\\
					 &=1-\exp\left[\frac{A_{co} \cdot B_{so}}{\ln(1-\beta_{so})}\left(\sum\limits_{l_k \in \Pi} d_{S_k, D_k}\right)^2\right]\nonumber\\
           &=1-\exp\left[\frac{\lambda_E \pi}{\ln(1-\beta_{so})}\left(\frac{\gamma_C}{\gamma_E }\right)^{\frac {2}{\alpha}}\left(\sum\limits_{l_k \in \Pi} d_{S_k, D_k}\right)^2\right].
\end{align*}
\end{IEEEproof}

We can see from Formula~(\ref{eq:optimal_P_COP}) that $P_{S_k}^{\text{SO-COP}}$ is an increasing function of $\lambda_J$, $P_{\bar{J}}$, $\gamma_E$ and $\beta_{so}$, while being a decreasing function of $\lambda_E$. We can see from Formula~(\ref{eq:optimal_COP}) that $P_{co}^*(\Pi)$ is an increasing function of $\gamma_C$ and $\lambda_E$, while being a decreasing function of $\gamma_E$ and $\beta_{so}$.

\subsection{QO-SOP: QoS-based Optimal SOP}

We then analyze how to achieve optimal secure performance (minimal SOP) conditioned on that QoS performance is ensured (COP is below some pre-specified threshold), which is termed as the problem QO-SOP.

Let $\beta_{co}$ ($0<\beta_{co}<1$) denote the pre-specified constraint on COP of path $\Pi$, then the problem QO-SOP can be formally defined as the following optimization problem:

\begin{eqnarray}
&\mathop {\min }\limits_{l_k \in \Pi,  P_{S_k} } & P_{so}(\Pi) \label{eq:QO-SOP} \\
&s.t. & P_{co}(\Pi)\leq \beta_{co}.  \label{eq:st-SOP}
\end{eqnarray}

Regarding the problem QO-SOP (\ref{eq:QO-SOP})-(\ref{eq:st-SOP}), we have the following theorem.

\begin{theorem} \label{theorem:QO-COP}
For a given multi-hop ad hoc network, where the densities of eavesdroppers and jammers are $\lambda_E$ and $\lambda_J$, respectively, the required SIRs for an intended receiver correctly decoding the message and an eavesdropper successfully intercepting the message are $\gamma_C$ and $\gamma_E$, respectively, the constraint on communication QoS is $\beta_{co}$, then the optimal solution (i.e., optimal transmission power) of problem QO-SOP is determined as:
\begin{equation}
\begin{aligned}
P_{S_k}^{\text{QO-COP}}  =\left[-\frac{A_{co}}{\ln(1-\beta_{co})} \cdot \left(\sum\limits_{l_k \in \Pi} d_{S_k,D_k}  \right) \cdot d_{S_k,D_k} \right]^{\alpha/2}, \\
           l_k \in \Pi,
\end{aligned}
\label{eq:optimal_P_SOP}
\end{equation}
and the optimal achievable SOP with the guaranteed COP is given by
\begin{equation}
P_{so}^*(\Pi)=1-\exp\left[\frac{\lambda_E \pi}{\ln(1-\beta_{co})}\left(\frac{\gamma_C}{\gamma_E}\right)^{\frac {2}{\alpha}}\left(\sum\limits_{l_k \in \Pi} d_{S_k, D_k}\right)^2\right]. \label{eq:optimal_SOP}
\end{equation}
\end{theorem}

\begin{IEEEproof}
Let $F_k=P_{S_k}^{2/\alpha}$, then $P_{co}(\Pi)$ and $P_{so}(\Pi)$ can be expressed as (\ref{eq:COP_1}) and (\ref{eq:SOP_1}), respectively. Substituting (\ref{eq:COP_1}) into (\ref{eq:st-SOP}) we have
\begin{equation}
\sum_{l_k \in \Pi}  \frac{d_{S_k, D_k}^2} {F_k} \leq  -\frac{\ln(1-\beta_{co})}{A_{co}} \triangleq \epsilon_{co}. \label{eq:F_k_SOP}
\end{equation}
Notice that $P_{so}(\Pi)$ in (\ref{eq:SOP_1}) is an increasing function of $F_k$ while the objective in (\ref{eq:QO-SOP}) is to minimize $P_{so}(\Pi)$, so the inequality constraint (\ref{eq:F_k_SOP}) can be replaced by the equality constraint $\sum\limits_{l_k \in \Pi} \frac{d_{S_k, D_k}^2} {F_k} = \epsilon_{co}$. Therefore, the problem QO-SOP is equivalent to the following optimization issue:
\begin{eqnarray}
&\mathop {\min }\limits_{l_k \in \Pi,  F_k } & \sum\limits_{l_k \in \Pi} {F_k} \label{eq:QO-SOP_1} \\
&s.t. &  \sum\limits_{l_k \in \Pi} \frac{d_{S_k,D_k}^2}{F_k} = \epsilon_{co}.  \label{eq:st-SOP_1}
\end{eqnarray}

Similar to the proof of Theorem~\ref{theorem:SO-COP}, we also apply the method of Lagrange multipliers and obtain the following $K$ equations:
\begin{equation}
\begin{aligned}
\frac{\partial}{\partial {F_k}}\left\{ \sum_{l_k \in \Pi} F_k + \theta_2 \left( \sum_{l_k \in \Pi} \frac{d_{S_k, D_k}^2}{F_k}- \epsilon_{co} \right) \right\}=0, \Bigg{|}_{F_k^*}=0, \\
l_k \in \Pi,
\end{aligned}
\end{equation}
where $\theta_2$ is the Lagrange multiplier. Then we have
\begin{align}
            & 1-\theta_2 \frac{d_{S_k, D_k}^2}{(F_k^*)^2}=0, \ l_k \in \Pi, \nonumber \\
\Rightarrow & F_k^*=\sqrt \theta_2 d_{S_k, D_k}, \ l_k \in \Pi. \label{eq:F_k_SOP_1}
\end{align}
Substituting (\ref{eq:F_k_SOP_1}) into (\ref{eq:st-SOP_1}), $\theta_2$ can be determined as:
\begin{equation}
\theta_2=\left(\frac{1}{\epsilon_{co}}\sum_{l_k \in \Pi} d_{S_k, D_k}\right)^2.
\label{eq:theta_2}
\end{equation}
Substituting (\ref{eq:theta_2}) into (\ref{eq:F_k_SOP_1}), we have
\begin{equation} \label{eq:opt_solve_co_cn_yk}
F_k^*=\frac{1}{\epsilon_{co}}\left(\sum\limits_{l_k \in \Pi} d_{S_k, D_k}\right) d_{S_k, D_k}, \ l_k \in \Pi.
\end{equation}
Thus, the optimal transmission power $P_{S_k}^{\text{QO-SOP}}$ of node $S_k$ is given by
\begin{equation*}
P_{S_k}^{\text{QO-SOP}}  =\left[-\frac{A_{co}}{\ln(1-\beta_{co})} \cdot \left(\sum\limits_{l_k \in \Pi} d_{S_k,D_k}  \right) \cdot d_{S_k,D_k} \right]^{\alpha/2},
\end{equation*}
and the minimum SOP $P_{so}^*(\Pi)$ of path $\Pi$ under the condition that $P_{co}(\Pi) \leq \beta_{co}$ is determined as:
\begin{align*}
P_{so}^*(\Pi) &=1-\exp\left(-B_{so}\sum_{l_k \in \Pi} F_k^* \right) \\
              &=1-\exp\left[-B_{so}\frac{1}{\epsilon_{co}}\left(\sum\limits_{l_k \in \Pi} d_{S_k, D_k}\right)^2\right] \\
							&=1-\exp\left[\frac{B_{so} \cdot A_{co}}{\ln(1-\beta_{so})}\left(\sum\limits_{l_k \in \Pi} d_{S_k, D_k}\right)^2\right] \\
              &=1-\exp\left[\frac{\lambda_E \pi}{\ln(1-\beta_{co})}\left(\frac{\gamma_C}{\gamma_E}\right)^{\frac {2}{\alpha}}\left(\sum\limits_{l_k \in \Pi} d_{S_k, D_k}\right)^2\right].
\end{align*}
\end{IEEEproof}

We can see from Formula~(\ref{eq:optimal_P_SOP}) that $P_{S_k}^{\text{QO-SOP}}$ is an increasing function of $\lambda_J$, $P_{\bar{J}}$ and $\gamma_C$, while being a decreasing function of $\beta_{co}$. We can see from Formula~(\ref{eq:optimal_SOP}) that $P_{so}^*(\Pi)$ is an increasing function of $\gamma_C$ and $\lambda_E$, while being a decreasing function of $\gamma_E$ and $\beta_{co}$. Furthermore, our results indicate that the jammer-related parameters $\lambda_J$ and $P_{\bar{J}}$ have impacts on $P_{co}(\Pi)$ and $P_{so}(\Pi)$, while have no impact on $P^*_{co}(\Pi)$ and $P_{so}^*(\Pi)$. This is in accordance with the intuition that jammers have opposite effects on COP and SOP, and for the performance tradeoffs the effects on two sides cancel each other out.

\section{Routing Algorithm} \label{section:routing}
In Section~\ref{section:outage_probabilities}, we have derived the expressions of outage probabilities for a given path, and in Section~\ref{section:optimization}, we have explored the optimal performance tradeoffs for a given path. Based on the obtained results, in this section, we further investigate the routing problem, i.e., for a pair of source and destination nodes with multiple optional end-to-end paths, how to select the optimal path to achieve the minimum COP under the security constraint or the minimum SOP under the QoS constraint.

\subsection{Routing Algorithm for SO-COP}
We first consider the routing algorithm for SO-COP. Based on Formula~(\ref{eq:optimal_COP}), the routing problem of finding the optimal path which achieves the minimum COP under the security constraint can be expressed as:
\begin{equation}
\min\limits_{\Pi \in \mathbf{S}(\Pi)} {1-\exp{\left[\frac{\lambda_E \pi}{\ln(1-\beta_{so})}\left(\frac{\gamma_C}{\gamma_E}\right)^{\frac {2}{\alpha}}\left(\sum\limits_{l_k \in \Pi} d_{S_k, D_k}\right)^2\right]}}, \label{eq:routing_SO-COP}
\end{equation}
where $\mathbf{S}(\Pi)$ denotes the set of all potential paths connecting the pair of source and destination nodes. Then (\ref{eq:routing_SO-COP}) is equivalent to
\begin{equation}
\min\limits_{\Pi \in \mathbf{S}(\Pi)} {\sum\limits_{l_k \in \Pi} d_{S_k, D_k}}. \label{eq:routing_SO-COP1}
\end{equation}

Expression~(\ref{eq:routing_SO-COP1}) indicates that the routing problem for SO-COP is equivalent to finding the shortest path connecting the pair of source and destination nodes. It means that we can assign the link weights $d_{S_k,D_k}$ to each potential link $l_k$ and then find the path $\Pi^*$ with the minimum total link weights. This problem can be directly solved by Bellman-Ford algorithm or Dijkstra's algorithm \cite{Medhi_BOOK10}, which returns the shortest paths from a source vertex to all other vertexes in a weighted graph. The computational complexity of Bellman-Ford algorithm is $O(N^3)$, while for Dijkstra's algorithm, it is $O(N^2)$, $N$ is the number of network nodes. However, Dijkstra's algorithm requires all link states in the whole network, while Bellman-Ford algorithm only needs the distance vectors between neighboring nodes such that it can be easily realized in a distributed ad hoc network based on the distance vector approach \cite{Medhi_BOOK10}.

The distributed Bellman-Ford algorithm (i.e., the distance vector approach) does not take security-QoS tradeoffs into consideration. Thus, after finding the shortest path $\Pi^*$, the routing algorithm for SO-COP should conduct the transmission power allocation for each node on path $\Pi^*$ (except the destination) based on Formula~(\ref{eq:optimal_P_COP}), which is another key procedure to achieve the optimal COP with a guaranteed SOP. It is notable that the computational complexity of proposed algorithm is dominated by the shortest path finding procedure, thus it has the same level of computational complexity as Bellman-Ford algorithm, i.e., $O(N^3)$. It is polynomial and much lower than that of the exhaustive search whose complexity is $O((N-2)!)$. The details of routing algorithm for SO-COP are summarized in Algorithm~\ref{algorithm:SO-COP}.

\begin{algorithm}[!ht]
\caption{Routing algorithm for SO-COP.}
\label{algorithm:SO-COP}
\begin{algorithmic}[1]
\REQUIRE
Network parameters $\{\lambda_J,\lambda_E,\gamma_C,\gamma_E,P_{\bar{J}},\alpha\}$ and security constraint $\beta_{so}$;
\ENSURE
The optimal path $\Pi^*$ for SO-COP, the corresponding transmission power $P_{S_k}^{\text{SO-COP}}$, the achievable COP $P_{co}^*(\Pi^*)$;
\STATE Initialization (assign the value $d_{S_k,D_k}$ to the link weights for any potential pair of transmitter $S_k$ and receiver $D_k$);
\STATE Find a shortest path in terms of the link weights between source node and destination node. The distributed Bellman-Ford algorithm can be applied for this procedure;
\STATE Assign the shortest path to $\Pi^*$;
\STATE Apply Formula~(\ref{eq:optimal_P_COP}) to allocate the corresponding transmission power $P_{S_k}^{\text{SO-COP}}$ for each transmitter on path $\Pi^*$;
\STATE Apply Formula~(\ref{eq:optimal_COP}) to calculate the secure-based optimal COP $P_{co}^*(\Pi^*)$ of path $\Pi^*$;
\RETURN $\{\Pi^*, P_{S_k}^{\text{SO-COP}}, P_{co}^*(\Pi^*)\}$;
\end{algorithmic}
\end{algorithm}

\subsection{Routing Algorithm for QO-SOP}
We then consider the routing algorithm for QO-SOP. Based on Formula~(\ref{eq:optimal_SOP}), the routing problem of finding the optimal path which achieves the minimum SOP under the QoS constraint can be expressed as:
\begin{equation}
\min\limits_{\Pi \in \mathbf{S}(\Pi)} {1-\exp{\left[\frac{\lambda_E \pi}{\ln(1-\beta_{co})}\left(\frac{\gamma_C}{\gamma_E}\right)^{\frac {2}{\alpha}}\left(\sum\limits_{l_k \in \Pi} d_{S_k, D_k}\right)^2\right]}}. \label{eq:routing_QO-SOP}
\end{equation}
Then (\ref{eq:routing_QO-SOP}) is equivalent to
\begin{equation}
\min\limits_{\Pi \in \mathbf{S}(\Pi)} {\sum\limits_{l_k \in \Pi} d_{S_k, D_k}}. \label{eq:routing_QO-SOP1}
\end{equation}

Expression~(\ref{eq:routing_QO-SOP1}) indicates that the routing problem for QO-SOP is also equivalent to finding the shortest path connecting the pair of source and destination nodes. Thus, we also apply the distributed Bellman-Ford algorithm to find the shortest path $\Pi^*$, and then allocate the transmission power of each node on path $\Pi^*$ (except the destination) based on Formula~(\ref{eq:optimal_P_SOP}). The details of routing algorithm for QO-SOP are summarized in Algorithm~\ref{algorithm:QO-SOP}.

\begin{algorithm}[!ht]
\caption{Routing algorithm for QO-SOP.}
\label{algorithm:QO-SOP}
\begin{algorithmic}[1]
\REQUIRE Network parameters $\{\lambda_J,\lambda_E,\gamma_C,\gamma_E,P_{\bar{J}},\alpha\}$ and QoS constraint $\beta_{co}$;
\ENSURE The optimal path $\Pi^*$ for QO-SOP, the corresponding transmission power $P_{S_k}^{\text{QO-SOP}}$, the achievable SOP $P_{so}^*(\Pi^*)$;
\STATE Initialization (assign the value $d_{S_k,D_k}$ to the link weights for any potential pair of transmitter $S_k$ and receiver $D_k$);
\STATE Find a shortest path in terms of the link weights between source node and destination node. The distributed Bellman-Ford algorithm can be applied for this procedure;
\STATE Assign the shortest path to $\Pi^*$;
\STATE Apply Formula~(\ref{eq:optimal_P_SOP}) to allocate the corresponding transmission power $P_{S_k}^{\text{QO-SOP}}$ for each transmitter on path $\Pi^*$;
\STATE Apply Formula~(\ref{eq:optimal_SOP}) to calculate the QoS-based optimal SOP $P_{so}^*(\Pi^*)$ of path $\Pi^*$;
\RETURN $\{\Pi^*, P_{S_k}^{\text{QO-SOP}}, P_{so}^*(\Pi^*)\}$;
\end{algorithmic}
\end{algorithm}

Regarding the implementation of the proposed routing algorithms for SO-COP and QO-SOP in a practical ad hoc network, our proposals could be incorporated in some existing distance vector routing protocols, either the proactive ones like M-DART \cite{Caleffi_WCMC11} or the reactive ones like AODV \cite{Perkins_03AODV}. The main difference between the two types of routings is that proactive routings establish and maintain routes proactively (periodically), while reactive routings construct and update routes only when needed (in an on-demand manner). Therefore, they should be applied based on network features, such as node mobility, delay requirement, etc. For example, for an ad hoc network with slow mobility and requires low delay, it is more appropriate for us to incorporate our proposals in proactive distance vector routing protocols, where the distributed Bellman-Ford algorithm can be applied to find the shortest path, and the dynamic addressing techniques in \cite{Caleffi_WCMC11} can be utilized to reduce routing overhead. For an ad hoc network with fast mobility and can tolerate large delay, the reactive protocols could be more efficient for the implementation of our proposals.  To deal with the fast topology changes caused by node mobility, route is established (i.e., finding the shortest path and allocating transmission power) in an on-demand manner. AODV is a variant of Bellman-Ford distance vector routing protocol, in which our proposals could be incorporated to achieve optimal security-QoS tradeoffs.

\section{Numerical Results and Discussions} \label{section:numerical_results}
In this section, we first present the Monte Carlo \cite{Binder_BOOK10} simulation results to validate our theoretical analysis for the outage probabilities in a concerned multi-hop ad hoc network, and then apply our theoretical results to illustrate the performance tradeoffs and the corresponding routing algorithms.

\subsection{Simulation Settings}
We simulate a multi-hop ad hoc network in a $2000\times 2000$ square area. The jammers (resp. eavesdroppers) are distributed at random positions which follow the homogeneous PPP with density $\lambda_J$ (resp. $\lambda_E$). Regarding the basic network parameters, we set $P_{\bar{J}}=1$, $\gamma_C=1$, $\gamma_E=1$ and $\alpha=4$. In each Monte Carlo simulation for COP and SOP, we consider the example of a fixed path $\Pi=\left\langle l_{1},\ldots,l_{5}\right\rangle$ with five links, where the transmission power $P_{S_k}$ and the distance $d_{S_k,D_k}$ of each link are set to be the same, respectively. The duration of each task of Monte Carlo simulation is set to $10^7$ rounds, and the simulated outage probability is given by
\begin{equation}
\text{simulated outage probability}=100\% \times \frac{N_o}{10^7}, \label{eq:Monte_Carlo}
\end{equation}
where $N_o$ denotes the number of times that the event of outage occurs in each simulation.

\subsection{Validation for COP and SOP}
\begin{figure}[!t]
\centering
\includegraphics[width=0.45\textwidth]{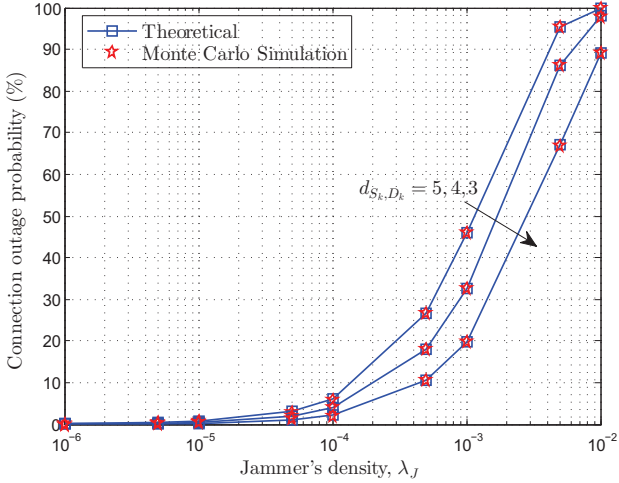}
\caption{Performance validation of COP: $P_{co}(\Pi)$ versus $\lambda_J$ under different settings of $d_{S_k,D_k}$. $K=5$, $P_{S_k}=1$ and $d_{S_k,D_k}=\{3,4,5\}$ for $1 \leq k \leq K $, $P_{\bar{J}}=1$, $\gamma_C=1$, $\alpha=4$.}
\label{fig:P_co}
\end{figure}

We first summarize in Fig.~\ref{fig:P_co} the theoretical and simulation results of COP performance, where we set $P_{S_k}=1$ and $d_{S_k,D_k}=\{3,4,5\}$ for $1\leq k \leq 5$. The theoretical curves are plotted according to Formula~(\ref{eq:COP}) while the simulated results are obtained based on Formula~(\ref{eq:Monte_Carlo}). We can see from Fig.~\ref{fig:P_co} that the simulation results match nicely with the theoretical ones for all the cases, which indicates that our theoretical analysis is highly efficient in the evaluation of end-to-end COP of multi-hop ad hoc networks. Another observation of Fig.~\ref{fig:P_co} is that as the jammer's density $\lambda_J$ and/or the transmission distance $d_{S_k,D_k}$ increase, COP increases and thus the communication QoS is degraded.

\begin{figure}[!t]
\centering
\includegraphics[width=0.45\textwidth]{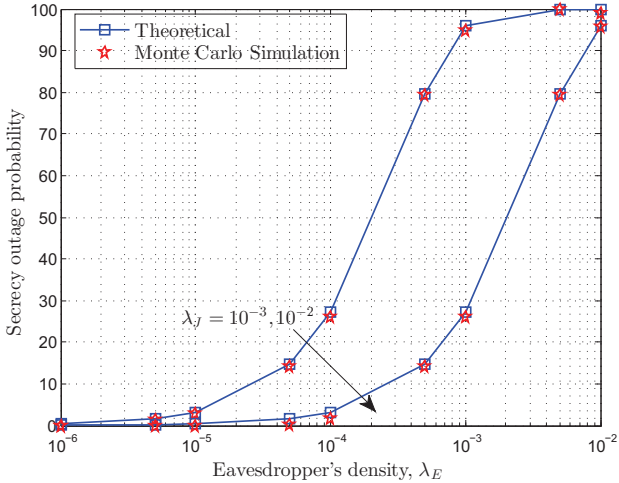}
\caption{Performance validation of SOP: $P_{so}(\Pi)$ versus $\lambda_E$ under different settings of $\lambda_J$. $K=5$, $P_{S_k}=1$ for $1 \leq k \leq K $, $\lambda_J=\{10^{-3},10^{-2}\}$, $P_{\bar{J}}=1$, $\gamma_E=1$, $\alpha=4$.}
\label{fig:P_so}
\end{figure}

We then summarize in Fig.~\ref{fig:P_so} the theoretical and simulation results of SOP performance, where we set $P_{S_k}=1$ for $1\leq k \leq 5$, and $\lambda_J=\{10^{-3},10^{-2}\}$. The theoretical curves are plotted according to Formula~(\ref{eq:SOP}) while the simulated results are obtained based on Formula~(\ref{eq:Monte_Carlo}). Similar to Fig.~\ref{fig:P_co}, Fig.~\ref{fig:P_so} shows that the simulation results match well with the theoretical ones for all the cases, which indicates that our theoretical analysis is highly efficient in the evaluation of end-to-end SOP of multi-hop ad hoc networks. We can also see from Fig.~\ref{fig:P_so} that SOP increases (thus the transmission security degrades) monotonically as the eavesdropper's density $\lambda_E$ increases, while increasing the jammer's density $\lambda_J$ will lead to a decrease in SOP, indicating that the jammers can be utilized cooperatively to improve the security performance.

\subsection{Performance Tradeoffs}

\begin{figure}[!t]
\centering
\includegraphics[width=0.45\textwidth]{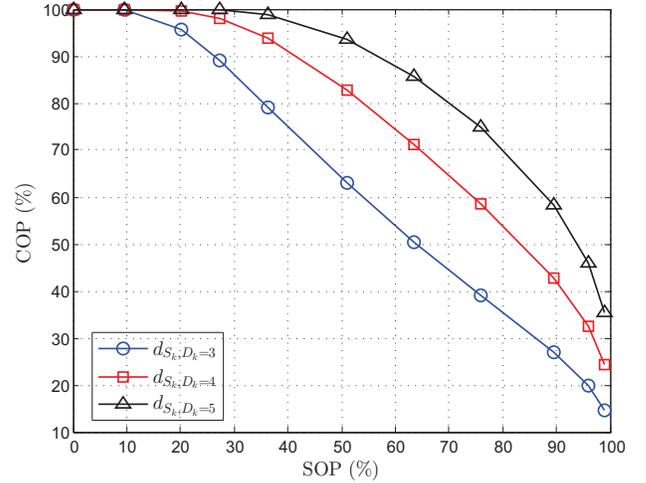}
\caption{COP-SOP tradeoff with the variation of transmission power. $K=5$, for the points from left to right on each curve, $P_{S_k}$ takes the value from the set $\{0,0.001,0.005,0.01,0.02,0.05,0.1,0.2,0.5,1,2\}$ sequentially, $\lambda_J=10^{-3}$, $\lambda_E=10^{-3}$, $\gamma_C=1$, $\gamma_E=1$, $P_{\bar{J}}=1$, $\alpha=4$.
}
\label{fig:tradeoff}
\end{figure}

We show in Fig.~\ref{fig:tradeoff} the COP-SOP tradeoff with the variation of transmission power, where $\lambda_J=10^{-3}$, $\lambda_E=10^{-3}$, and we consider a path $\Pi$ with five links (i.e., $K=5$), each of which has the same distance and same power. For the points from left to right on each curve of Fig.~\ref{fig:tradeoff}, the transmission power $P_{S_k}$ takes the value from the set $\{0,0.001,0.005,0.01,0.02,0.05,0.1,0.2,0.5,1,2\}$ sequentially. We can see from Fig.~\ref{fig:tradeoff} that as $P_{S_k}$ increases, SOP increases while COP decreases, indicating that the tradeoffs between transmission security and communication QoS can be achieved by controlling the transmission power. A further careful observation of Fig.~\ref{fig:tradeoff} is that for the same SOP (for example, $P_{so}=50\%$), the minimum $d_{S_k,D_k}$ can lead to the minimum COP ($P_{co}$ is $64\%$, $83\%$ and $94\%$ under $d_{S_k,D_k}=\{3,4,5\}$, respectively), which indicates that a shorter transmission distance can lead to a better performance tradeoff.

\begin{figure}[H]
    \centering
		\subfigure[$P_{co}^*(\Pi)$ versus $\beta_{so}$ ($P_{so}^*(\Pi)$ versus $\beta_{co}$).]
    {\includegraphics[width=0.45\textwidth]{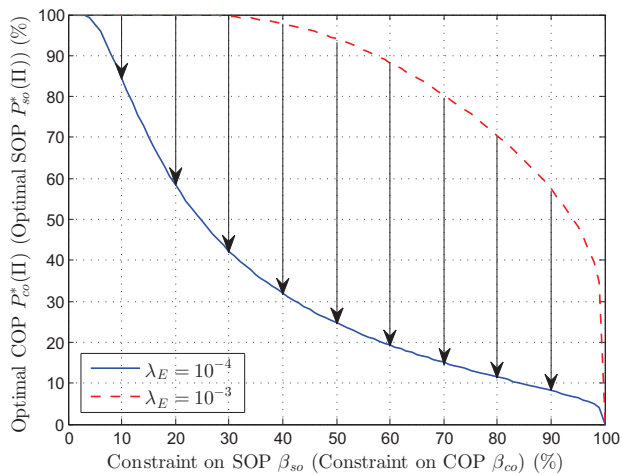} \label{fig:P_beta}}
		\hfill
    \subfigure[$P_{S_k}^{\text{SO-COP}}$ versus $\beta_{so}$. $\lambda_J=10^{-3}$.]
    {\includegraphics[width=0.45\textwidth]{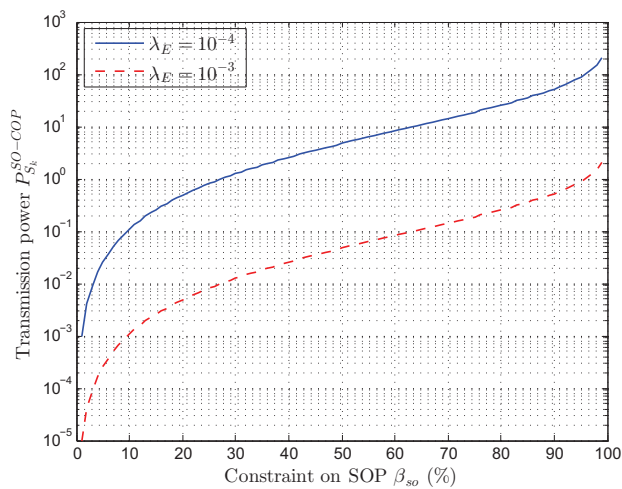} \label{fig:P_SOCOP}}
   	\hfill
		\subfigure[$P_{S_k}^{\text{QO-SOP}}$ versus $\beta_{co}$.]
    {\includegraphics[width=0.45\textwidth]{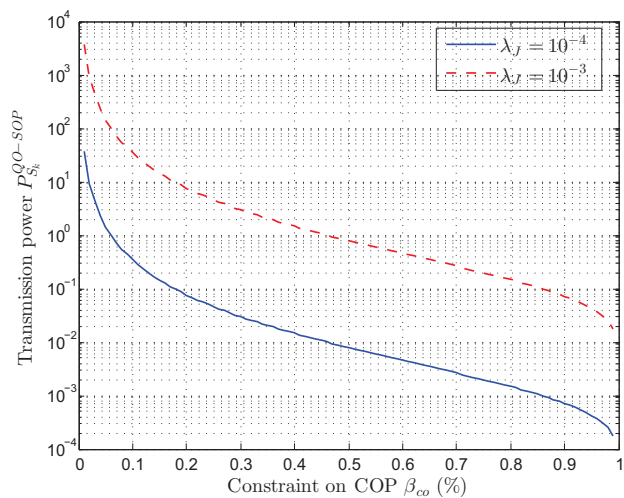} \label{fig:P_QOSOP}}
		\caption{Performance of SO-COP and QO-SOP. $K=5$, $d_{S_k,D_k}=5$ for $1 \leq k \leq K$, $P_{\bar{J}}=1$, $\gamma_C=1$, $\gamma_E=1$, $\alpha=4$.}
		\label{fig:optimal_tradeoff}
\end{figure}

We summarize in Fig.~\ref{fig:optimal_tradeoff} the performance of SO-COP and QO-SOP, where we set $K=5$ and $d_{S_k,D_k}=5$ for $1 \leq k \leq K$. It is worth noting that the expressions of $P_{co}^*(\Pi)$ and $P_{so}^*(\Pi)$ are almost the same, except that $\beta_{co}$ in (\ref{eq:optimal_COP}) is replaced by $\beta_{so}$ in (\ref{eq:optimal_SOP}). Thus, we plot Fig.~\ref{fig:P_beta} to show how $P_{co}^*(\Pi)$ varies with $\beta_{so}$ and how $P_{so}^*(\Pi)$ varies with $\beta_{co}$, simultaneously. We can see from Fig.~\ref{fig:P_beta} that as $\beta_{so}$ (resp. $\beta_{co}$) increases, which means the constraint on SOP (resp. COP) declines, $P_{co}^*(\Pi)$ (resp. $P_{so}^*(\Pi)$) decreases monotonically. Another observation about Fig.~\ref{fig:P_beta} is that a big gap exists between the curves under $\lambda_E=10^{-4}$ and $\lambda_E=10^{-3}$, which indicates the eavesdropper's density has a great impact on the network performance.

Fig.~\ref{fig:P_SOCOP} shows how the transmission power $P_{S_k}^{\text{SO-COP}}$ to achieve SO-COP varies with $\beta_{so}$, and Fig.~\ref{fig:P_QOSOP} shows how the transmission power $P_{S_k}^{\text{QO-SOP}}$ to achieve QO-SOP varies with $\beta_{co}$. We can see that $P_{S_k}^{\text{SO-COP}}$ increases monotonically as $\beta_{so}$ increases, while $P_{S_k}^{\text{QO-SOP}}$ decreases monotonically as $\beta_{co}$ increases. Moreover, Fig.~\ref{fig:P_SOCOP} indicates that to deal with the network scenario with denser eavesdroppers, for example, increasing $\lambda_E$ from $10^{-4}$ to $10^{-3}$, we should diminish the transmission power; while Fig.~\ref{fig:P_QOSOP} indicates that to deal with the network scenario with denser jammers, for example, increasing $\lambda_J$ from $10^{-4}$ to $10^{-3}$, we should increase the transmission power.

Since we set the distance of each link on path $\Pi$ is the same in Fig.~\ref{fig:optimal_tradeoff}, the corresponding transmission power of each link is also the same. To further illustrate the performance under the network scenario with different link length, we consider a path $\Pi$ which consists of five links, the distance of each link is uniformly distributed on $(1,10)$. We set $\lambda_J=10^{-3}$, $\lambda_E=10^{-4}$, $\beta_{so}=0.5$ and $\beta_{co}=0.5$, and the results of one implementation are summarized in Table~\ref{table:tradeoff}.

\begin{table}[!ht]
\centering
\caption{the transmission power of each link and the optimal outage probabilities.}
\renewcommand\arraystretch{1.8}
\begin{tabular}{|c|c|c|c|c|c|}
\hline
The $k^{\text{th}}$ link & 1 & 2 & 3 & 4 & 5 \\
\hline
$d_{S_k,D_k}$ &   3.5726  &  7.8148 &   7.7836 &   4.4240 &   6.1104 \\
\hline
$P_{S_k}^{\text{SO-COP}}$ &  1.7147  &  8.2046  &  8.1391  &  2.6294  &  5.0160 \\
\hline
$P_{S_k}^{\text{QO-SOP}}$ & 0.5708   & 2.7314   & 2.7097   & 0.8754  &  1.6699 \\
\hline
$P_{co}^*(\Pi)$, $P_{so}^*(\Pi)$ & \multicolumn{5}{c|}{0.3269} \\
\hline
\end{tabular}
\label{table:tradeoff}
\end{table}

\subsection{Routing Performance}

\begin{figure}[!t]
\centering
\includegraphics[width=0.45\textwidth]{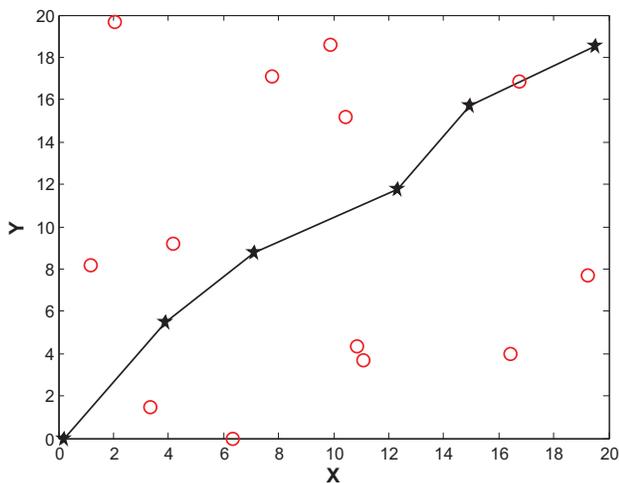}
\caption{A snapshot of the optimal path for SO-COP and QO-SOP. The optimal path $\Pi^*$ connecting the source and destination are plotted by the black solid line with ``$\bigstar$'', and other legitimate nodes are plotted by red empty circles.}
\label{fig:routing}
\end{figure}

To illustrate the routing algorithms for SO-COP and QO-SOP, we focus on a $20 \times 20$ square area and randomly place $20$ legitimate nodes following the uniform distribution. We assign the node which is closest to the lower left corner as the source, and assign the node which is closest to the upper right corner as the destination. Notice that the eavesdroppers and jammers are still randomly distributed over the whole network area, and we set the densities as $\lambda_E=10^{-4}$ and $\lambda_J=10^{-3}$. In order to ensure the end-to-end transmission is formed by multiple hops, we strategically set the maximal transmission range of a single hop as $8$.

We plot in Fig.~\ref{fig:routing} a snapshot of the optimal path for SO-COP and QO-SOP. For the snapshot of network scenario in Fig.~\ref{fig:routing}, the optimal path $\Pi^*$ with the shortest path length is selected by executing the Bellman-Ford algorithm. Based on the distance of each link on path $\Pi^*$, our proposed routing algorithms allocate the transmission power for each link to achieve the optimal performance tradeoffs. Here we set both $\beta_{so}$ and $\beta_{co}$ as $0.4$, then the optimal achievable COP and SOP, as well as the corresponding transmission power of each link are summarized in Table~\ref{table:routing}.

\begin{table}[H]
\centering
\caption{link length, transmission power and optimal outage probabilities of the optimal path $\Pi^*$.}
\renewcommand\arraystretch{1.8}
\begin{tabular}{|c|c|c|c|c|c|}
\hline
The $k^{\text{th}}$ link & 1 & 2 & 3 & 4 & 5 \\
\hline
$d_{S_k,D_k}$  &  6.6027  &  4.6456  &  5.9676  &  4.7477  &  5.3562 \\
\hline
$P_{S_k}^{\text{SO-COP}}$ &  3.7608  &  1.8617  &  3.0721  &  1.9444  &  2.4748 \\
\hline
$P_{S_k}^{\text{QO-SOP}}$ &  3.0366  &  1.5033  &  2.4806  &  1.5700  &  1.9983 \\
\hline
$P_{co}^*(\Pi^*)$, $P_{so}^*(\Pi^*)$ & \multicolumn{5}{c|}{0.3681} \\
\hline
\end{tabular}
\label{table:routing}
\end{table}

\section{Conclusion and Future Work} \label{section:conclusion}
This paper studied the physical layer security-aware routing and the performance tradeoffs between transmission security and communication QoS in multi-hop ad hoc networks. Considering a multi-hop ad hoc network which consists of randomly distributed legitimate nodes, cooperative jammers and malicious eavesdroppers, we first derived the closed-form expressions of COP and SOP for a given path. Then, we analyzed the security-QoS tradeoffs to obtain the minimum achievable COP (resp. SOP) with a guaranteed SOP (resp. COP) and the corresponding strategies of power allocation for each transmitter on the path. With the help of theoretical analysis of a given path, we finally proposed the Bellman-Ford based routing algorithms to find the optimal path between any pair of source and destination nodes which can achieve the optimal security-QoS tradeoffs.

Notice that this work is based on the interference-limited assumption and didn't consider the combining techniques at the eavesdroppers, so one promising future direction is to extend our study to the more realistic scenarios with the considerations of additive noises and combining techniques. Another appealing future work is to explore the physical layer security-aware routing in a more general ad hoc network where jammers have different transmission power.

\section*{Acknowledgment}
This work was partially supported by the Project of Cyber Security Establishment with Inter-University Cooperation, Secom Science and Technology Foundation, Japan JSPS Grant 15H02692, China NSFC Grants 61571352, 61373173 and U1536202, Research Foundation for Youths 20103176192, Recruitment Program of Foreign Experts MS2016XADZ046.

\end{document}